\documentclass[showpacs, showkeys,superscriptaddress]{revtex4}
\usepackage{bm,showlabels}

\usepackage{tikz}
\usetikzlibrary{matrix}
\usetikzlibrary{calc}
\usepackage{rotating}
\usepackage{amssymb,amsmath,amsthm,cancel,color}
\usepackage{titlesec}
\usepackage{hyperref}
\usepackage[applemac]{inputenc}
\allowdisplaybreaks

\titleformat{\paragraph}
{\normalfont\normalsize\bfseries}{\theparagraph}{1em}{}
\titlespacing*{\paragraph}
{0pt}{3.25ex plus 1ex minus .2ex}{1.5ex plus .2ex}

\begin{document}

\title{Constraining Generalized Non-local Cosmology from Noether Symmetries}

\author{Sebastian Bahamonde}
\email{sebastian.beltran.14@ucl.ac.uk}
\affiliation{Department of Mathematics, University College London,
	Gower Street, London, WC1E 6BT, United Kingdom}

\author{Salvatore Capozziello}
\email{capozziello@na.infn.it}
\affiliation{Dipartimento di Fisica "E. Pancini", Universit\'a di Napoli
	\textquotedblleft{Federico II}\textquotedblright, Napoli, Italy,}
\affiliation{Gran Sasso Science Institute, Via F. Crispi 7, I-67100, L' Aquila,
	Italy,}
\affiliation{INFN Sez. di Napoli, Compl. Univ. di Monte S. Angelo, Edificio G, Via
	Cinthia, I-80126,
	Napoli, Italy.}
\author{Konstantinos F.	Dialektopoulos}
\email{dialektopoulos@na.infn.it}
\affiliation{Dipartimento di Fisica "E. Pancini", Universit\'a di Napoli
	\textquotedblleft{Federico II}\textquotedblright, Napoli, Italy,}
\affiliation{INFN Sez. di Napoli, Compl. Univ. di Monte S. Angelo, Edificio G, Via
	Cinthia, I-80126,
	Napoli, Italy.}
\begin{abstract}
\textbf{Abstract:} We study a generalized nonlocal theory of gravity which, in specific limits, can become either the  curvature non-local  or teleparallel non-local theory. Using the Noether Symmetry Approach,  we find that  the coupling functions coming from the non-local terms are constrained to be either exponential or linear in form. It is well known that in some non-local theories, a certain kind of exponential non-local couplings are needed in order to achieve a renormalizable theory.  In this paper,  we explicitly show that this kind of coupling does not need to by introduced by hand, instead, it appears naturally from the symmetries of the Lagrangian in flat Friedmann-Robertson-Walker cosmology. Finally, we  find de-Sitter and power law cosmological solutions for different nonlocal theories. The symmetries for the generalized non-local theory is also found and some cosmological solutions are also achieved under the full theory. 
\end{abstract}


\date{\today}

\maketitle
	
\section{Introduction}
Apart from its remarkable success to interprete cosmological observations, the $\Lambda$-Cold Dark Matter ($\Lambda$CDM) model still lacks in according a satisfactory explanation to the issue why the energy density of the cosmological constant is so small if compared to the vacuum energy of the Standard Model (SM) of particle physics.  Furthermore, the today observed equivalence, in order of magnitude, of dark matter and dark energy escapes any  general explanation a part the introduction of a very strict fine tuning. 

Starting from these facts,  one cannot consider the cosmological constant fully responsible for the whole anti-gravity dynamics, like the incapability to find a convincing candidate for dark matter, or/and a quantum theory of gravity,   many scientists started questioning whether the theory, i.e. General Relativity (GR), needed to be changed, in order to explain the accelerating expansion and the large scale structure clustering without the introduction of  {\it "ad hoc"}  cosmological constant and new particles, see, for example, \cite{Capozziello:2011et,reportsergei,Clifton:2011jh} and many others. The most usual modifications consists in the introduction of new fields either in the matter sector (e.g. quintessence) or by modifying gravity (e.g. scalar-tensor theories, $f(R)$, $f(T)$, etc.). In some sense, the issue is related to adding new matter fields (dark matter, quintessence, etc.) or improving the geometry considering further degrees of freedom of gravitational field.

Almost a decade ago, a non-local modification of the Einstein-Hilbert (EH) action has been proposed \cite{Deser:2007jk}, and the new action has the following form
 \begin{eqnarray}
 	\mathcal{S}_{\rm standard-NL}&=&\frac{1}{2\kappa}\int d^{4}x\, \sqrt{-g(x)} \, R(x)\left[1 + f\Big((\square^{-1}R)(x)\Big) \right]+\int d^{4}x\, \sqrt{-g(x)}\,L_{m}\,, \label{1b}
 \end{eqnarray}	 
 where $\kappa=8\pi G$, $R$ is the Ricci scalar,  $f$ is an arbitrary function which depends on the retarded Green function evaluated at the Ricci scalar, $L_{m}$ is any matter Lagrangian and  $\square \equiv \partial_{\rho}(e g^{\sigma\rho}\partial_{\sigma})/e$ is the 
 scalar-wave operator, which can be written in terms of the  Green function $G(x,x')$ as 
 \begin{eqnarray}
 	(\square^{-1}F)(x)=\int d^4x'\, e(x') F(x')G(x,x')\,.\label{G}
 \end{eqnarray}
 It is clear that by setting $f(\square^{-1}R)=0$, the above action is equivalent to the Einstein-Hilbert one plus the matter content.  The non-locality is introduced by the inverse of the d'Alembert operator (see \cite{Deser:2007jk} for details). Corrections of this kind arise naturally as soon as quantum loop effects are studied and they are also considered as possible solution to the black hole information paradox \cite{Donoghue:1994dn,Giddings:2006sj}. Since then, a lot of studies of non-localities have been done in various contexts \cite{univ4,modesto1,modesto2,st1,st2,loop,jm}. 
 In Refs. \cite{Mod1,Mod2,Mod3,Mod4,Mod5}, non local quantum gravity is fully discussed putting in evidence results and open issues.  From the string  theory point of view, in \cite{Arefeva:2007wvo} they present some bouncing solutions, in \cite{Arefeva:2007xdy} solutions of an expanding Universe with phantom dark energy and in \cite{Barnaby:2008fk} they generate non-Gaussianities during inflation. Emanating from infrared (IR) scales, a lot of progress has also been done. Unification of inflation with late-time acceleration, as well as, the dynamics of a local form of the theory have been studied in \cite{Nojiri:2007uq,Jhingan:2008ym}. In \cite{Deser:2013uya}, they show that non-local gravity models do not alter the GR predictions for gravitationally bound systems, and also they are ghost-free and stable. Finally, in \cite{Deffayet:2009ca,Koivisto:2008dh,Koivisto:2008xfa}, they derived a technique to fix the functional form of the function $f$ in the action, which is called nonlocal distortion function. The interested reader should see the detailed review by Barvinsky \cite{Barvinsky:2014lja}, which summarizes the non-local aspects both from the quantum-field theory point of view and from the cosmological one.

Along another track, teleparallel \cite{Pereira} and modified teleparallel theories of gravity \cite{Cai:2015emx,manos} have, in the last decade, gained a lot of attention trying not only to formulate gravity in a gauge invariant way, but also to interpret the late-time acceleration of the Universe, without invoking any {''ad hoc"}  cosmological constant. The idea is that gravity, instead of curvature, is mediated only through torsion. This means that, the theory is no more a geometrical theory, i.e. the trajectories of the particles are not described by geodesic equations, but just by some force equations, since torsion is seen as a force, similar to the Lorentz equation in electrodynamics. The Teleparallel Equivalent of General Relativity (TEGR) is a gauge description of the gravitational interactions and torsion  defined through the Weitzenb\"ock connection (instead of the Levi-Civita connection, used by GR, where the Equivalence Principle is strictly requested in order to make geodesic and metric structure to coincide). Hence, in this theory, the manifold is flat but endorsed with torsion.  The dynamical fields of the theory are the four linearly independent vierbeins and their relation with the metric and the inverse of the metric is given by
\begin{equation}
g_{\mu\nu} = \eta _{ab}e^a_{\mu} e^b_{\nu}\,, \quad g^{\mu\nu} = \eta^{ab}E_a^{\mu} E_b^{\nu}\,,
\end{equation}
where $\eta _{ab}$ is the flat Minkowski metric and $E_a {}^{\mu} $ is the inverse of the tetrads. The action of TEGR is given by 
\begin{equation}
\mathcal{S}_{\rm TEGR} =-\frac{1}{2\kappa} \int d^4 x e\,T  +\int d^{4}x\, e\,L_{m}\,,
\end{equation}
with $e$ being $e= \det(e^i{}_{\mu}) = \sqrt{-g}$ and $T$ is the torsion scalar, which is given by the contraction
\begin{equation}
T = S^{\mu\nu}{}_{\rho} T^{\rho}{}_{\mu\nu}\,,
\end{equation}
where
\begin{eqnarray}
S_{\rho}{}^{\mu\nu} &=& \frac{1}{2}\left(K^{\mu\nu}{}_{\rho}+\delta^{\mu}{}_{\rho}T^{\sigma \nu}{}_{\sigma}-\delta^{\nu}{}_{\rho}T^{\sigma \mu}{}_{\sigma}\right)\,, \\
K^{\mu\nu}{}_{\rho} &=& -\frac{1}{2}\left(T^{\mu\nu}{}_{\rho} - T^{\nu\mu}{}_{\rho} - T_{\rho}{}^{\mu\nu} \right)\,,  \\
T^{\alpha}{}_{\mu\nu} &=& \Gamma^{\alpha}{}_{\mu\nu} - \bar{\Gamma}^{\alpha} {}_{\mu\nu}\,,
\end{eqnarray}
are respectively the superpotential, the contorsion tensor, the torsion tensor and $\bar{\Gamma}^{\alpha}{}_{\mu\nu}=E_{a}^{\alpha}\partial_{\mu}e_{\nu}^{a}$ is the Weitzeb\"ock connection. The teleparallelism condition gives the relation of the Ricci scalar with the torsion scalar, that is 
\begin{equation}\label{RT}
R = -T + \frac{2}{e}\partial_{\mu}(e T^{\mu}) = -T + B\,.
\end{equation}
Hence, we directly see that at the action level, the EH action with the TEGR action differ only by a boundary term and thus the descriptions are equivalent. This is easily generalized to a more complex action as soon as   we substitute $T$ with an arbitrary function of this, $f(T)$. This theory can present problems being  non-Lorentz invariant and because a covariant formulation of $f(T)$ gravity  is still not very well accepted since the spin connection is a  field without dynamics. Nevertheless, it is always possible to  give rise to the correct field equations  choosing suitable  tetrads (see the review paper \cite{Cai:2015emx} for a detailed discussion on advantages and problems related to $f(T)$ gravity). 

The extra degrees of freedom introduced by $f$, do not allow us to find an exact  relation between $f(T)$ and $f(R)$, since now the boundary terms in \eqref{RT}, contribute to the field equations. These kind of theories and their extensions   are of great interest \cite{Capozziello:2016eaz,Paliathanasis:2016vsw,Bahamonde:2016kba,Chen:2010va,Li:2011wu}, since they provide theoretical interpretation of the accelerating expansion of the Universe and also accomodates the radiation and matter dominated phases of it. In specific cases, one can also find inflationary solutions and avoid the Big Bang singularity with bouncing solutions. \\
In the teleparallel framework, recently it was proposed a similar kind of non-local gravity based on the torsion scalar $T$. In this theory, the action reads as follows \cite{our}
\begin{eqnarray}
\mathcal{S}_{\rm teleparalell-NL}&=&-\frac{1}{2\kappa}\int d^{4}x\, e(x) \, T(x)+\frac{1}{2\kappa}\int d^{4}x\, e(x) \, T(x)f\Big((\square^{-1}T)(x)\Big)+\int d^{4}x\, e(x)\,L_{m}\,, \label{1}
\end{eqnarray}	 
where $e=\textrm{det}(e^{a}_{\mu})=\sqrt{-g}$ and now the function $f$ depends on $\square^{-1}T$. The teleparallel equivalent of GR is recovered if $f(\square^{-1}T)=0$. It is possible to show  \cite{our} that this theory is consistent with the cosmological data by SNe Ia + BAO + CC +$H_0$  observations. From (\ref{RT}), it is straightforward  noticing that (\ref{1b}) and (\ref{1}) correspond to different theories, where $B$ is the  term  connecting them.

Let us now present a generalization of (\ref{1b}) and (\ref{1}), which we call Generalized Non-local Teleparallel Gravity (GNTG). Its action is given by 
\begin{equation}
\label{ourtheory}
 \mathcal{S} =  -\frac{1}{2\kappa}\int d^{4}x\, e T +\frac{1}{2\kappa}\int d^{4}x\, e(x) \, (\xi T(x)+\chi B(x))f\Big((\square^{-1}T)(x),(\square^{-1}B)(x)\Big)+\int d^{4}x\, e(x)\,L_{m}\,. 
\end{equation}
Here, $T$ is the torsion scalar, $B$ is a boundary term and $f(\square ^{-1}T,\square^{-1}B)$ is now an arbitrary function of the nonlocal torsion and the nonlocal boundary terms. The greek letters $\xi$ and $\chi$ denote coupling  constants. It is easily seen, that by choosing $\xi= -\chi=-1$ one obtains the standard Ricci scalar. From (\ref{G}), we directly see that the following relation is also true
\begin{equation}
\square^{-1}R = -\square ^{-1}T + \square ^{-1}B\,,
\end{equation}
and thus, if $f(\square ^{-1}T,\square^{-1}B) = f(-\square ^{-1}T + \square^{-1}B)$, the action takes the well known form $Rf(\square ^{-1}R)$ given by the action (\ref{1b}). Moreover, nonlocal teleparalell gravity given by the action (\ref{1})  is recovered if $\chi=0$ and $f(\square^{-1}T,\square^{-1}B)=f(\square^{-1}T)$. Starting from this theory, we can construct a scalar tensor analog by using Lagrange multipliers and we can constrain the distortion function $f$ by the so-called  Noether Symmetries Approach \cite{cimento}. There is a huge amount of articles in the literature, which adopt the Noether Symmetry Approach to constrain  the form of some classes  theories (see for example \cite{Capozziello:2011et, Capozziello:2016eaz, Bahamonde:2016grb} and references therein). In this way, one obtains models that, thanks to the existence of Noether Symmetries, present integrals of motion that allows to reduce dynamics and then, in principle,  to find out exact solutions. Besides these technical points, the presence of symmetries fixes couplings and potentials with physical meaning \cite{cimento}. In such a way, the approach can be considered a sort of criterion to "select" physically motivated theories \cite{NoetherQC}. Details on the approach  will be given in Sec. \ref{noetherS}.

The paper is organised as follows: in Sec. \ref{sec:1} we present details on the model, how to construct the action and its scalar-tensor analog with four auxiliary fields. At the end of this section, we present a diagram which shows the different theories that we can construct as subclasses of the general  theory. In the next Sec. \ref{noetherS}, we summarize the Noether Symmetry Approach that, we shall apply to three different cases: i) the teleparallel non-local case (a coupling like $Tf(\square ^{-1}T)$), in the Section \ref{telenonlocal}; ii) the curvature nonlocal gravity (a coupling like $Rf(\square ^{-1}R)$), in the Section \ref{standnonlocal}; and iii) the generalized non-local case (given by the complete action \eqref{ourtheory}), in the Section \ref{gennonlocal}.
In each case, after the symmetries study, we present a set of cosmological solutions. 
 Discussion and conclusions are reported in \eqref{conclusions}.  The Appendix \ref{Appendix} is devoted to details on the conditions to select the Noether vector. Throughout the paper we adopt the signature $(+,-,-,-)$.

\section{Generalized Non-local Cosmology }\label{sec:1}
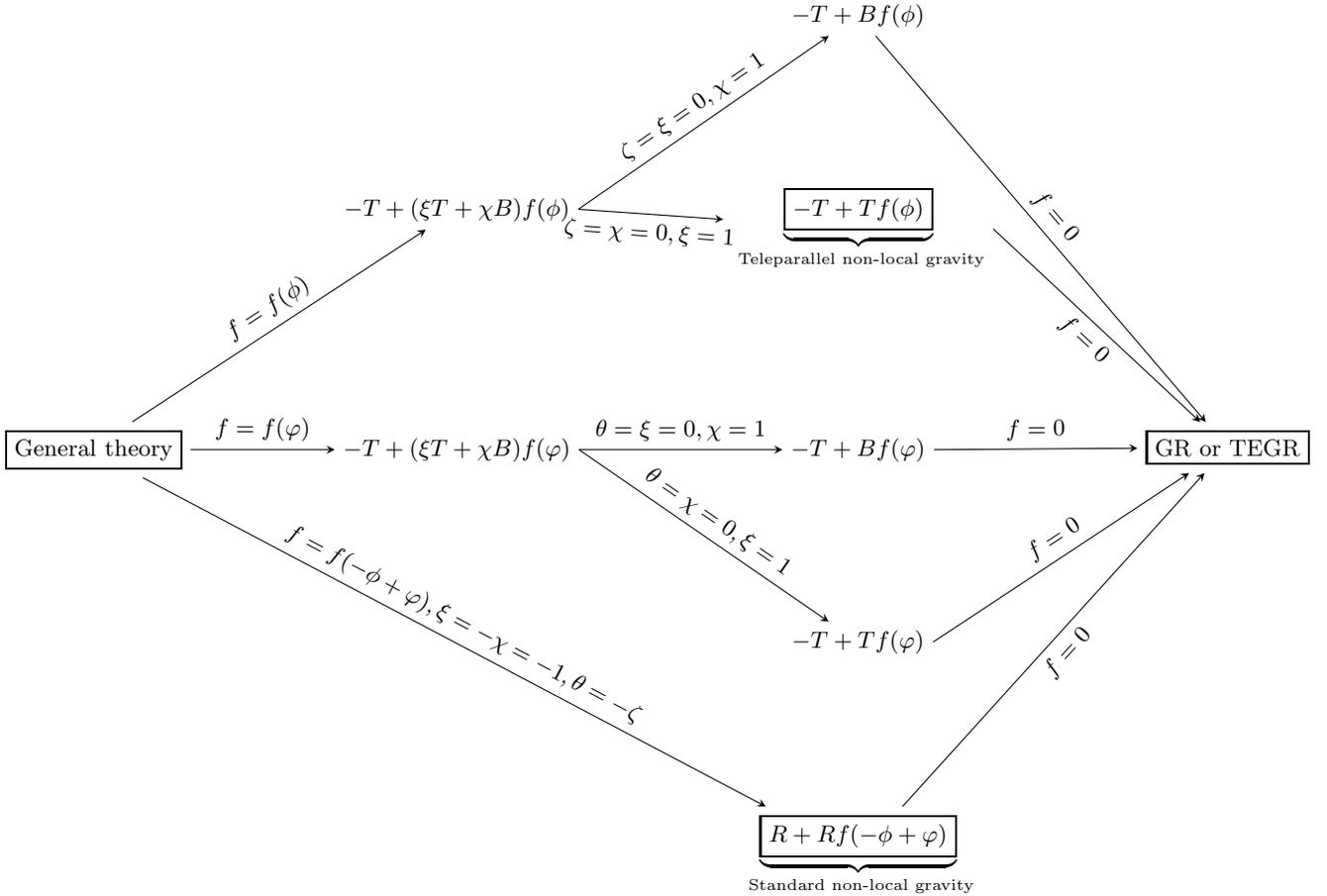
\begin{figure}[!ht]
	\centering
	\begin{tikzpicture}
	\matrix (m) [matrix of math nodes,row sep=6em,column sep=6em,minimum width=2em]
	{  & \ & -T+Bf(\phi)& \\\ 
		\ & -T+(\xi T+\chi B)f(\phi) & \underbrace{\boxed{-T+Tf(\phi)}}_{\textrm{ Teleparallel non-local gravity}}& \\\
		\boxed{\textrm{General theory}} &  -T+(\xi T+\chi B)f(\varphi) & -T+Bf(\varphi)& \boxed{\text{GR or TEGR}} \\
		\ & \ & -T+Tf(\varphi)\\
		\ & \ & \underbrace{\boxed{R+Rf(-\phi+\varphi)}}_{\textrm{ Standard non-local gravity}}& \ \\
	};
	\path[-stealth,every node/.style={sloped,anchor=south}]
	(m-3-1) edge node  {$f =f(\phi)$} (m-2-2)
	(m-2-2.east|-m-2-2) edge node  {$\zeta=\xi=0,\chi=1$} (m-1-3)
	(m-3-1.east|-m-3-1) edge node [above] {$f=f(\varphi)$} (m-3-2)
	(m-3-2.east|-m-3-2) edge node [above] {$\theta=\xi=0,\chi=1$} (m-3-3)
	(m-3-3.east|-m-3-3) edge node [above] {$f=0$} (m-3-4)
	(m-3-2.east|-m-3-2) edge node  {$\ \theta=\chi=0,\xi=1$} (m-4-3)
	(m-4-3.east|-m-4-3) edge node  {$\ f=0$} (m-3-4)
	(m-1-3) edge node  {$f=0$} (m-3-4)
	(m-2-2.east|-m-2-2) edge node[below]  {$ \zeta=\chi=0,\xi=1$} (m-2-3)
		(m-2-3.east|-m-2-3) edge node[below]  {$ f=0$} (m-3-4)
	(m-3-1) edge node  {$f=f(-\phi+\varphi),\xi=-\chi=-1,\theta=-\zeta$}(m-5-3)
	(m-5-3) edge node [below] {$f=0$} (m-3-4);
	\end{tikzpicture}
	\caption{The diagram shows how to recover the different theories of gravity starting from the scalar-field representation of the general theory. Note that $\phi=\square^{-1}T$ and $\varphi=\square^{-1}B$ so that $-\phi+\varphi=\square^{-1}R$. Clearly, the curvature and torsion representaions "converge" only for the linear theories in $R$, the GR, and in $T$, the TEGR.}
		\label{fig1}
\end{figure}

Since the field equations for the GNTG theory are very cumbersome, we will rerewrite the action (\ref{ourtheory}) in a more suitable way using scalar fields, according to  \cite{Nojiri:2007uq}. Specifically, the action can be rewritten introducing four scalar fields $\phi,\psi,\theta,\zeta$ as follows
\begin{eqnarray}
	\mathcal{S}&=& -\frac{1}{2\kappa}\int d^{4}x\, e T +  \frac{1}{2\kappa}\int d^{4}x\, e\left[ \, (\xi T+\chi B)f(\phi,\varphi) +\theta (\square \phi - T)+  \zeta (\square \varphi -B)\right]+ \int d^{4}x\, e\,L_{m} \,,\nonumber  \\
	&=& -\frac{1}{2\kappa}\int d^{4}x\, e T + \frac{1}{2\kappa}\int d^{4}x\, e\left[ \, (\xi T+\chi B)f(\phi,\varphi) -  \partial_{\mu} \theta \partial^{\mu} \phi - \theta T -  \partial _{\mu} \zeta \partial ^{\mu} \varphi -\zeta B \right]+ \int d^{4}x\, e\,L_{m}\,.  \label{action2}
\end{eqnarray}
By varying this action with respect to $\theta$ and $\zeta$ we get $\phi= \square^{-1}T$ and $\varphi= \square^{-1}B$ respectively. In addition, by varying this action with respect to $\phi$ and $\varphi$ we get
\begin{eqnarray}
 \square \theta= (\xi T + \chi B) \frac{\partial f(\phi,\varphi)}{\partial\phi}\,,\\
\square \zeta= (\xi T + \chi B) \frac{\partial f(\phi,\varphi)}{\partial\varphi} 	\,.
\end{eqnarray}
In the scalar representation is not  straightforward  how  recovering curvature or teleparallel nonlocal gravity. Let us explicitly recover these theories under scalar formalism.  For example, by setting $\xi = -1 = - \chi$,  $f(\phi,\varphi) = f(-\phi+\varphi)$, and $\theta =- \zeta$ we obtain standard non-local curvature gravity, namely
\begin{eqnarray}
\mathcal{S}_{\rm standard-NL}&=& \frac{1}{2\kappa}\int d^{4}x\, \sqrt{-g} \left[ \,R + R f(\psi) - \partial_{\mu} \zeta \partial^{\mu} \psi -  \zeta R  \right]+ \int d^{4}x\, e\,L_{m}\,,\\
&=& \frac{1}{2\kappa}\int d^{4}x\, \sqrt{-g} \left[ \, R + R f(\square^{-1}R)  \right]+ \int d^{4}x\, e\,L_{m}\,, \label{curvaction}
\end{eqnarray}
where $\psi = -\phi + \varphi$. On the other hand,  the non-local TEGR is recovered if in the action \eqref{action2} we choose $\xi = 1\,,\,\chi=0$, $f(\phi,\varphi) = f(\phi)$ and $\zeta=0$. We obtain
\begin{eqnarray}
\mathcal{S}_{\rm teleparallel-NL}&=& \frac{1}{2\kappa}\int d^{4}x\, e\left[ \, T\left(f(\phi)-1\right) - \partial_{\mu} \theta \partial^{\mu} \phi -  \theta T  \right]+ \int d^{4}x\, e\,L_{m}\label{teleaction0} \\
&=& \frac{1}{2\kappa}\int d^{4}x\, e\left[ \, T\left(f(\square^{-1}T)-1\right)  \right]+ \int d^{4}x\, e\,L_{m}\,.  \label{teleaction}
\end{eqnarray}
A more general class of theories, like $-T + (\xi T+\chi B) f(\square ^{-1}T)$ or $-T + (\xi T+\chi B) f(\square ^{-1}B)$ can be obtained by setting $f(\phi,\varphi) = f(\phi)$ and $f(\phi,\varphi) = f(\varphi)$ respectively. Obviously, in these cases, one can change the values of $\xi$ and $\chi$ to obtain other  couplings like 
\begin{eqnarray}
S&=&\frac{1}{2\kappa}\int d^{4}x\, e\left[ \, -T + B f(\square^{-1}T)  \right]+ \int d^{4}x\, e\,L_{m}\,,
 \\ 
 S&=&\frac{1}{2\kappa}\int d^{4}x\, e\left[ \,-T + T f(\square^{-1}B) \right]+ \int d^{4}x\, e\,L_{m}\,, \\
 \  S&=&  \frac{1}{2\kappa}\int d^{4}x\, e\left[ -T + B f(\square^{-1}B) \right] + \int d^{4}x\, e\,L_{m}\,.
\end{eqnarray}
Fig.~\ref{fig1} is a comprehensive diagram representing  all the  theories that  can be  recovered from the action (\ref{action2}). Here, we have not considered unnatural couplings like $Rf(\square^{-1}T)$ or $Tf(\square^{-1}R)$ because $R$ and $T,B$ are quantities defined in different connections, so mixed terms like $Rf(\square^{-1}T)$ are badly defined. The above half part of the figure represents different non-local teleparallel theories and the below part of it, the standard curvature counterpart. As it is easy to see, only TEGR and GR dynamically coincide while this is not the case for other theories defined by $T$, $R$ and $B$. From a  fundamental point of view, this fact is extremely relevant because the various representations of gravity can have different dynamical contents. For example, it is well known that $f(T)$ gravity gives second order field equations while $f(R)$ gravity, in metric representation,  is fourth order. These facts are strictly related to the dynamical roles of torsion and curvature and their discrimination at fundamental level could constitute an important insight to really understand the nature of gravitational field (see \cite{Cai:2015emx} for a detailed discussion).

By varying the generalized non-local action (\ref{action2}) with respect to the tetrads, we get the following field equations
\begin{eqnarray}
2(1-\xi (f(\phi,\varphi)-\theta))\left[ e^{-1}\partial_\mu (e S_{a}{}^{\mu\beta})-E_{a}^{\lambda}T^{\rho}{}_{\mu\lambda}S_{\rho}{}^{\beta\mu}-\frac{1}{4}E^{\beta}_{a}T\right]- \nonumber\\
-\frac{1}{2}\Big[(\partial^{\lambda}\theta)(\partial_{\lambda}\phi)E_{a}^{\beta}-(\partial^{\beta}\theta)(\partial_{a}\phi)-(\partial_{a}\theta)(\partial^{\beta}\phi)\Big]-\frac{1}{2}\Big[(\partial^{\lambda}\zeta)(\partial_{\lambda}\varphi)E_{a}^{\beta}-(\partial^{\beta}\zeta)(\partial_{a}\varphi)-(\partial_{a}\zeta)(\partial^{\beta}\varphi)\Big]+ \nonumber\\ 
+2\,\partial_{\mu}\Big[f(\phi,\varphi)(\xi+\chi)-\theta-\zeta\Big]E^\rho_a S_{\rho}{}^{\mu\nu} + \Big(E^{\nu}_{a}\Box-E^\mu_a \nabla^{\nu}\nabla_{\mu}\Big)(\zeta-\chi f(\phi,\varphi))=  \kappa\Theta^\beta_a\,, \label{eq22}
\end{eqnarray}
where $\Theta^\beta_a$ is the general energy-momentum tensor.\\
Let us now take into account  the tetrad $
e^{a}_{\beta}=(1,a(t),a(t),a(t)), \,\label{tetrad}
$ which reproduces the flat  Friedmann-Robertson-Walker (FRW) metric   $ds^2=dt^2-a(t)^2(dx^2+dy^2+dz^2)$. For this geometry, the modified FRW equations are
\begin{gather}
3 H^2 (\theta-\xi f+1) = \frac{1}{2}   \dot{\zeta} \dot{\varphi}+\frac{1}{2}  \dot{\theta}\dot{\phi}+3 H   \big(\dot{\zeta}-\chi\dot{f}\big)+\kappa \rho_m\,,\\
\big(2 \dot{H}+3 H^2\big) (\theta-\xi  f+1) = -  \frac{1}{2}\dot{\zeta} \dot{\varphi}-\frac{1}{2}  \dot{\theta} \dot{\phi}-\dot{f} (2 H (\xi +2 \chi )+\chi )
+2H(2 \dot{\zeta}+\dot{\theta})+  \ddot{\zeta}-\kappa p_m\,,
\end{gather}
where $\rho_m$ and $p_m$ are the energy density and the pressure of the cosmic fluid respectively and dots denote differentiation with respect to the cosmic time. The equations for the scalar fields can be written as 
\begin{eqnarray}
6 H^2+3 H \dot{\phi}+\ddot{\phi}&=&0\,,\label{phieq}\\
6 (\dot{H}+3 H^2)+3 H \dot{\varphi}+\ddot{\varphi}&=&0\,,\label{varphieq}\\
-6 H^2 \left(\xi  f_{\varphi}+3 \chi  f_{\varphi}\right)-6 \dot{H} \chi  f_{\varphi}+3 H   \dot{\zeta}+ \ddot{\zeta}&=&0\,,\label{zetaeq}\\
-6 H^2 \left(\xi  f_{\phi}+3 \chi  f_{\phi}\right)-6 \dot{H} \chi  f_{\phi}+3 H  \dot{\theta}+   \ddot{\theta}&=&0\,,\label{thetaeq}
\end{eqnarray}
where the sub-indices represent the partial derivative $f_{\phi}=\partial f/\partial \phi$ and $f_{\varphi}=\partial f/\partial \varphi$. In the following section, we will use the Noether Symmetry Approach to seek for conserved quantities.

\section{The Noether Symmetry Approach}\label{noetherS}
Let us use the Noether Symmetry Approach \cite{cimento,Paliathanasis:2015mxa} in order to find symmetries and cosmological solutions for the generalized action (\ref{action2}). For simplicity, hereafter we will study the vacuum case, i.e., $\rho_m=p_m=0$. It can be shown that the torsion scalar and the boundary term in a flat FRW are given by
\begin{eqnarray}
T=-6H^2\,, \ \ B=-18H^2-6\dot{H}\,,
\end{eqnarray}
so that the action (\ref{action2}) takes the following form
\begin{equation}
\mathcal{S}=2\pi ^{2}\int a^3dt\left\{ -6\frac{\dot{a}^2}{a^2}(\xi f(\phi,\varphi) -\theta - 1)-6\Big(2\frac{\dot{a}^2}{a^2} - \frac{\ddot{a}}{a}\Big)(\chi f(\phi,\varphi) - \zeta)-\dot{\theta}\dot{\phi}-\dot{\zeta}\dot{\varphi}\right\} \,. \label{actioncan2}
\end{equation}
Considering the procedure in \cite{cimento}, we  find that the point-like Lagrangian is given by
\begin{eqnarray}\label{generalizedlagra}
\mathcal{L}=6 a\dot{a}^2 \big(\theta+1-\xi f(\phi,\varphi)\big)+6 a^2\dot{a}(\chi\dot{f}(\phi,\varphi)-\dot{\zeta})- a^3 \dot{\theta}\dot{\phi}- a^3 \dot{\zeta}\dot{\varphi}\,.
\end{eqnarray} 
The generator of infinitesimal transformations \cite{Paliathanasis:2015mxa} is given by
\begin{equation}
X = \lambda (t,x^{\mu})\partial_t + \eta ^i (t,x^{\mu})\partial _i\,,
\end{equation}
where $x^{\mu}=(a,\theta,\phi,\varphi,\zeta)$ and the vector $\eta^i$ is
\begin{align}
\eta ^i (t,x^{\mu})=\Big(\eta^{a},\eta^{\theta},\eta^{\phi},\eta^{\varphi},\eta^{\zeta}\Big)\,.
\end{align}
In general, each function depends on $t$ and $x^{\mu}$. If there exists a function $h=h(t,x^{\mu})$ such that
\begin{equation}\label{symcond}
X^{[1]}\mathcal{L}+\mathcal{L}\frac{d\lambda}{dt} = 
\frac{dh}{dt}\,,
\end{equation}
where $\mathcal{L}=\mathcal{L}(t,x^{\mu},\dot{x}^{\mu})$ is the Lagrangian of a system and $X^{[1]}$ is the first prolongation of the vector $X$ \cite{Paliathanasis:2015mxa}, then the Euler-Lagrange equations remain invariant under these transformations. The generator is a Noether symmetry of the system described by $\mathcal{L}$ and the relative integral of motion is given by
\begin{equation}
I = \lambda \left( \dot{x}^{\mu}\frac{\partial \mathcal{L}}{\partial \dot{x}^{\mu}}-\mathcal{L}\right) - \eta ^i \frac{\partial \mathcal{L}}{\partial \dot{x}^{\mu}} + h\,.
\end{equation}

In the next subsections, we will search for Noether symmetries in specific non-local Lagrangians, starting from the two  cases $(Tf(\square^{-1}T) \,\text{and}\, Rf(\square^{-1}R))$ and ending up to  the general action (\ref{action2}). 
The set of generalized coordinates $x^{\mu} = \{t,a,\theta,\phi,\varphi,\zeta\}$ gives rise to the  configuration space $\mathcal{Q} \equiv \{x^{\mu}\,,\mu=1,...,6\}$  and the tangent space  $\mathcal{T}\mathcal{Q} \equiv \{x^{\mu},\dot{x}^{\mu}\}$ of the Lagrangian
$\mathcal{L}=\mathcal{L}(t,x^{\mu},\dot{x}^{\mu})$.
Clearly, the procedure can be applied to   many different models starting from  Fig.~\ref{fig1}.

\section{Noether's symmetries in teleparallel non-local gravity with coupling $Tf(\square^{-1}T)$}
\label{telenonlocal}
\subsection{Finding Noether's symmetries}
Let us first study the case where we recover the teleparalel non-local case studied in \cite{our}. In this case, the torsion scalar $T$ is coupled with a non-local function evaluated at the torsion scalar, that is  $f(\square^{-1}T)=f(\phi)$. For Noether's symmetries, we   need to consider,
\begin{eqnarray}
f(\phi,\varphi)=f(\phi)\,, \quad  \chi=0\,, \quad \xi=1 \,\quad \text{and} \,\quad \zeta = 0\,.
\end{eqnarray}
in the general action \eqref{action2} and thus the Lagrangian becomes
\begin{equation}\label{lagratele}
\mathcal{L} = 6 a \left(- f(\phi) + \theta  + 1\right) \dot{a}^2 - a^3 \dot{\theta} \dot{\phi}\,.
\end{equation}
From Eq. \eqref{symcond},  one derives a system of 16 equations for the coefficients of the Noether vector and the functions $h,f$. It can be immediately seen that the dependence on the coordinates of the Noether vector components is 
\begin{eqnarray}
\lambda(a,\theta,\phi,t) &=& \lambda(t)\,,\\
\eta_{a}(a,\theta,\phi,t) &=& \eta_{a}(a,\theta,\phi,t)\,,\\
\eta_{\phi}(a,\theta,\phi,t) &=& \eta_{\phi}(a,\phi,t)\,,\\
\eta_{\theta}(a,\theta,\phi,t) &=& \eta_{\theta}(a,\theta,t)\,,\\
h(a,\theta,\phi,t) &=& h(a,\theta,\phi)\,.
\end{eqnarray}  
The whole system can be  straightforwardly derived from the general one in Appendix \ref{Appendix} (see also  \cite{Paliathanasis:2015mxa, Basilakos} for details). Note that we do not need to impose any ansantz to find out the symmetries. Hence, the  equation for $f$ reads
\begin{equation}\label{eq1}
c_1 f'(\phi) - c_2 f(\phi) + c_2 - c_3 =0 \,,
\end{equation}
where $c_1,c_2$ and $c_3$ are constants. There are two non trivial solutions ($f\neq$ constant)  to \eqref{eq1} depending on the value of $c_2$, i.e.
\begin{equation}\label{ftele}
f(\phi)=
\left \{ \begin{array}{rcl}
c_7 e^{\frac{c_2 \phi }{c_1}}-\frac{c_3}{c_2}+1\,, && c_{2}\neq 0\,,\\
c_7 + \frac{c_3 }{c_1}\phi \,, && c_2 = 0\,,
\end{array}\right.
\end{equation}
where $c_7$ is another integration constant.  From (\ref{teleaction}), we can notice that for having a TEGR (or GR) background we must have that $c_3=c_2$ in the exponential form and $c_7=0$ in the linear form. The Noether vector has the following form
\begin{equation}
X = (c_4 + c_5 t) \partial_t -\frac{1}{3}(c_2-c_4)a\partial _a +( c_3+c_2 \theta) \partial _{\theta} + c_1 \partial _{\phi} \,,
\end{equation}
and the integral of motion is 
\begin{eqnarray}
I = a^3 c_1 \dot{\theta} + a^3 c_2 (\theta +1) \dot{\phi} - a^3 \left(c_4 t+c_5\right) \dot{\theta} \dot{\phi} +\left[4 a^2 \left(c_2-c_4\right) \dot{a}+6 a \dot{a}^2 \left(c_4 t+c_5\right) \right] (-f(\phi )+\theta +1)+c_6\,.
\end{eqnarray}

\subsection{Cosmological solutions}
In the previous subsection we found that the form of the function $f$ is constrained to be an exponential or a linear form of the non-local term \eqref{ftele}. It can be shown that for the linear form, there are no power-law or de-Sitter solution. Here we will  find solutions for the exponential form of the coupling function.\\
As we pointed out before, it is physically convenient to choose $c_2=c_3$ in order to have a GR (or TEGR) background.  Hence, in this section,  we will assume this condition for the constants. 	For the exponential form of the function $f(\phi)$ given by \eqref{ftele}, the Lagrangian \eqref{lagratele} takes now the form
\begin{equation}
\mathcal{L} =  -6 a\dot{a}^2 \left(c_7 e^{\frac{c_3 \phi }{c_1}}-\theta -1\right)- a^3 \dot{\theta} \dot{\phi}\,,
\end{equation}
so that the Euler-Lagrange equations are given by
\begin{gather}\label{eqa}
c_1 \left(4 \dot{H} (c_7 e^{\frac{c_2 \phi}{c_1}}-\theta-1)-\dot{\theta} \dot{\phi}\right)+H \left(4 c_2 c_7 \dot{\phi} e^{\frac{c_2 \phi}{c_1}}-4 c_1 \dot{\theta}\right)+6 c_1 H^2 \left(c_7 e^{\frac{c_2 \phi}{c_1}}-\theta-1\right)= 0 \,,\\ \label{eqtheta}
6 H^2+3 H \dot{\phi}+\ddot{\phi} = 0 \,,\\ \label{eqphi}
-\frac{6 c_2 c_7 }{c_1}H^2 e^{\frac{c_2 \phi}{c_1}}+\ddot{\theta}+3 H \dot{\theta} = 0 \,, \\ \label{eneq}
6H^2 \left(- c_7 e^{\frac{c_2 \phi}{c_1}}+ \theta+1\right) - \dot{\theta} \dot{\phi}+6 \theta  H^2 = 0 \,,
\end{gather}
for $a,\theta,\phi$ and the energy equation, respectively. If we consider de-Sitter solution for the scale factor, $$a(t) = e^{H_0 t} \Rightarrow H(t) = H_0\,,$$ we immediately find from \eqref{eqtheta} that
\begin{equation}
\phi (t) = -2 H_0 t-\frac{\phi _1 e^{-3 H_0 t}}{3 H_0} + \phi _2 \,.
\end{equation}
For the sake of simplicity, we will choose $\phi _1 =\phi_2= 0$ otherwise Eq. \eqref{eqphi} cannot be integrated easily. By this assumption,  we directly find that
\begin{equation}
\theta (t) = e^{-3 H_0 t} \left(-c_7 (3 H_0 t+1)-\frac{\theta_1}{3 H_0}\right)+\theta_2\,,
\end{equation}
where $\theta_1$ and $\theta_2$ are integration constants and we needed to choose the branch $c_1=2c_2/3$, otherwise Eq. (\ref{eqa}) cannot be satisfied. Hence, from (\ref{eqa}) we directly see that $\theta_2=-1$, giving us the following cosmological solution,
\begin{equation}
a(t) = e^{H_0 t} \,,\quad \phi (t) = -2 H_0 t \,, \quad \theta (t) = e^{-3 H_0 t} \left(-c_7 (3 H_0 t+1)-\frac{\theta_1}{3 H_0}\right)-1\,,
\end{equation}
and 
\begin{equation}
f(\phi) = c_7 e^{-3 H_0 t}\,.
\end{equation}
If we consider that the scale factor behaves as a power-law $a(t)=a_0t^p$, where $p$ is a constant, from (\ref{eqtheta}) we directly find that 
\begin{eqnarray}
\phi(t)=\frac{6 p^2 \log (t-3 p t)}{1-3 p}+\frac{\phi_1}{1-3p} t^{1-3 p}+\phi_0\,,
\end{eqnarray}
where $\phi_1$ and $\phi_0$ are integration constants that for simplicity (as we did before) we will assume that are zero, otherwise (\ref{eqphi}) cannot be integrated directly. By doing this, we find
\begin{eqnarray}
\theta(t)=\frac{c_1 t^{1-3 p}}{1-3 p}+c_2+\frac{c_7 (3 p-1) (c_1-3 c_1 p) }{c_1 (1-3 p)^2-6 c_2 p^2}(t-3 p t)^{\frac{6c_2 p^2}{c_1-3 c_1 p}}\,,
\end{eqnarray}
where $\theta_0$ and $\theta_1$ are integration constants and we have assumed that $c_1\neq\tfrac{6 c_2 p^2}{(3 p-1)^2}$ and $p\neq1/3$ since there are not solutions for these other two branches. By replacing this solution into (\ref{eqa}) we get that $c_2 = \frac{c_1 (2 - 9 p + 9 p^2)}{6 p^2}$ and $\theta_1=-1$ yielding the following solution
\begin{eqnarray}
\phi(t)=\frac{6 p^2 \log (t-3 p t)}{1-3 p}\,,\quad \theta(t)=c_7 (1-3p)^{3-3p} t^{2-3p}+\frac{\theta_0 t^{1-3 p}}{1-3 p}-1\,, \quad a(t)=a_0t^p\,, \quad f(\phi)=c_7e^{\frac{\left(9 p^2-9 p+2\right) \phi }{6 p^2}}\,.
\end{eqnarray}
Note that  the energy condition (\ref{eneq}) is satisfied and $p=1/3$ is not a solution.

\section{Noether's symmetries in curvature non-local gravity with coupling $Rf(\square^{-1}R)$}
\label{standnonlocal}
\subsection{Finding Noether's symmetries}
Let us  find now Noether's symmetries for the case where curvature non-local gravity  is  considered.  We assume that the coupling  $Rf(\square^{-1}R)$ is present in the action. To recover this case, we must set
\begin{eqnarray}
f(\phi,\varphi)=f(-\phi+\varphi)=f(\psi)\,,\quad  \chi=1\,, \quad \xi=-1\,,\quad \theta = - \zeta\,.
\end{eqnarray}
In this way, the Lagrangian \eqref{action2} reads as follows
\begin{equation}\label{standlagra}
\mathcal{L} =  6  a\dot{a}^2 (f(\psi)+\theta + 1) +6 a^2  \dot{a}(f'(\psi )\dot{\psi}+ \dot{\theta})  + a^3 \dot{\theta} \dot{\psi}\,.
\end{equation}
and   Noether's condition equation \eqref{symcond},  gives a system of 18 differential equations. Also this is a special case of that presented in Appendix \ref{Appendix}.  The  result is 
\begin{equation}
\lambda (a,\theta,\psi,t) = \lambda (t)\, \quad \text{and} \quad h(a,\theta,\psi,t) = h(a,\theta,\psi)\,,
\end{equation}
and the system reduces to 9 equations. However, the full system is still difficult to be solved without any assumption. A simple assumption is choosing  $h(a,\theta,\psi)$ = constant.  
The last two equations of Noether condition for $f(\psi)$ are
\begin{eqnarray}
2 c_2 f'(\psi )+c_1 f(\psi )+c_1-c_3 &=& 0\,,\label{52}\\
2 c_2 f''(\psi )+c_1 f'(\psi ) &=& 0 \,.\label{53}
\end{eqnarray}
and the Noether vector results to be
\begin{eqnarray}
X=(c_5 + c_4 t)\partial_t + \frac{1}{3} a (c_4 - c_1)\partial_{a} + (c_3 + c_1 \theta )\partial_{\theta} -2 c_2 \partial_{\psi} \,.
\end{eqnarray}
Eqs. \eqref{52} and \eqref{53} are easily solved and the form of $f$ is
\begin{equation}\label{eq2}
f(\psi)=
\left \{ \begin{array}{rcl}
-1+\frac{c_3}{c_1} + c_6 e^{-\frac{c_1}{c_2} \psi} && c_1 \neq 0\,,\\
c_6 + \frac{c_3 }{2 c_2}\psi \,. && c_1 = 0
\end{array}\,.\right.
\end{equation}
Again, the form of the function  is  either exponential or linear in $\psi=\square^{-1}R$. This result is very interesting since, without further assumptions than   $h=const.$, the symmetries give the same  kind of couplings for both teleparallel and curvature non-local theories. These two couplings can be particularly relevant to get a renormalizable theory of gravity. As discussed in \cite{ModestoQ1,ModestoQ2}, the form of the coupling is extremely important to achieve a regular theory. In particular, the exponential coupling plays an important role in calculations. Here, the  symmetry itself is imposing this kind of coupling. In other words,   it  is not put  by hand but  is related to a fundamental principle, i.e. the existence of the Noether symmetry.

\subsection{Cosmological solutions}
It is well known  \cite{Nojiri:2007uq} that, non-local theories with  exponential coupling, i.e. $R(1+ e^{\alpha \square^{-1}R})$, have both de-Sitter and power-law solutions. In this section, we will verify that the Lagrangian \eqref{standlagra} with the coupling \eqref{eq2}, given by the symmetry, i.e.
\begin{equation}
\mathcal{L} = 6 a  \left( \frac{c_3}{c_1} + \theta \right) \dot{a}^2 + 3 c_6  a e^{-\frac{c_1}{2 c_2} \psi } \left(2 \dot{a}^2 - \frac{c_1}{c_2} a \dot{a}\dot{\psi}\right) + 6 a^2 \dot{a} \dot{\theta} + a^3 \dot{\theta} \dot{\psi}\,,
\end{equation}
gives rise to these solutions. In order to recover the  GR background, we will assume that $c_3=c_1$.\\
Let us start from the de-Sitter case, where $a(t) = e^{H_0 t}$. The Euler-Lagrange equations for $a, \psi, \theta$ and the energy equation, read respectively
\begin{gather}
c_1^2 c_6\dot{\psi}^2+8 c_2^2 \dot{H} \left(\theta e^{\frac{c_1 \psi}{2 c_2}}+e^{\frac{c_1 \psi}{2 c_2}}+c_6\right)+12 c_2^2 H^2 \left(\theta e^{\frac{c_1 \psi}{2 c_2}}+e^{\frac{c_1 \psi}{2 c_2}}+c_6\right)\nonumber\\
+4 c_2^2 \ddot{\theta} e^{\frac{c_1 \psi}{2 c_2}}-2 c_2^2 \dot{\theta}\dot{\psi} e^{\frac{c_1 \psi}{2 c_2}}+4 c_2 H \left(2 c_2 \dot{\theta} e^{\frac{c_1 \psi}{2 c_2}}-c_1 c_6\dot{\psi}\right)-2 c_1 c_2 c_6 \ddot{\psi}= 0 \,,\label{eqagr}\\ \label{eqpsi}
 3 c_1 c_6 \dot{H} e^{-\frac{c_1 \psi}{2 c_2}}+6 c_1 c_6 H^2 e^{-\frac{c_1 \psi}{2 c_2}}-3 c_2 H \dot{\theta}-c_2 \ddot{\theta} = 0 \,,\\ \label{eqthetagr}
 6 \dot{H}+3 H\dot{\psi}+12 H^2+\ddot{\psi}= 0 \,,\\ \label{eneqgr}
H \left(6 \dot{\theta}-\frac{3 c_1 c_6\dot{\psi} e^{-\frac{c_1 \psi}{2 c_2}}}{c_2}\right)+6 H^2 \left(c_6 e^{-\frac{c_1 \psi}{2 c_2}}+\theta+1\right)+\dot{\theta}\dot{\psi} =0 \,.
\end{gather}
Eq. \eqref{eqthetagr}  gives
\begin{equation}
\psi (t) = -4 H_0 t -\frac{\psi _1 e^{-3 H_0 t}}{3 H_0} + \psi _2\,,
\end{equation}
where $\psi_1$ and $\psi_0$ are integration constants. For simplicity, to find analytical solutions, we  set $\psi_1=\psi_0 =0$. Then, from Eq. \eqref{eqpsi} we find 
\begin{equation}
\theta (t) = \frac{3 c_2 c_6 }{2 c_1+3 c_2} e^{\frac{4 c_1 H_0 t - c_1 \psi _2}{2 c_2}} -\frac{\theta _1 }{3 H_0} e^{-3 H_0 t}+ \theta _2\,,
\end{equation}
and, in order to satisfy the other two Eqs. \eqref{eqagr} and \eqref{eneqgr}, we set $\theta _2 = - 1$ and $ c_2 = -c_1$.  Finally, the following  de-Sitter solution,
\begin{equation}
a(t) = e^{H_0 t} \,, \quad \psi(t) = - 4 H_0 t + \psi _2 \,, \quad \theta(t) = 3 c_6 e^{\frac{\psi _2}{2}-2 H_0 t}-\frac{\theta _1}{3 H_0} e^{-3 H_0 t}-1\,,
\end{equation}
is recovered and
\begin{equation}
f(\psi) = c_6 e^{\psi/2}\,.
\end{equation}
In the same spirit, if we assume that the scale factor with a power-law behavior as $a(t) = a_0t^p $, the system \eqref{eqagr}-\eqref{eneqgr} yields the following solution,
\begin{equation}
a(t) =a_0 t^p  \,, \quad \psi(t) = \frac{6 p (1-2 p)}{3 p-1}  \ln (t) \,, \quad \theta(t) = \frac{c_6 (3 p-1)}{(p-1) }t^{-2 p}-1 \,, \quad f(\psi(t)) =   c_6 e^{\frac{\psi (1-3p)}{3(1-2p)}}\,.
\end{equation}
This solution is  valid for $p\neq 1/3$.
Now, if one considers the linear form of $f(\psi)=c_6+\frac{c_3}{2c_2}\psi$, it is also possible to find power-law solutions but only for $p=1/2$, which corresponds to radiation. The non-trivial solution for this particular case is given by
\begin{eqnarray}
\theta(t)=\theta_0\,,\quad a(t)=a_0 t^{1/2}\,, \quad \psi(t)=-\frac{2 c_2 (\theta_0+2)}{c_3}-2\psi_1 t^{-1/2}\,, \quad f(\psi)=\frac{c_3 \psi }{2 c_2}+c_6\,,
\end{eqnarray}
where $\theta_0$ and $\psi_1$ are constants. 

\section{Noether's symmetries in the general case}\label{gennonlocal}
\subsection{Finding Noether's symmetries}
Let us consider now the generalized  non-local action involving both teleparallel and curvature non-local contributions.  The  Lagrangian is
\begin{equation}\label{genlagra}
\mathcal{L} = 6 \chi  a^2 \dot{a} \dot{\phi} f_{\phi}(\phi ,\varphi)+6 \chi  a^2 \dot{a} \dot{\varphi} f_{\varphi}(\phi ,\varphi )-6 \xi  a \dot{a}^2 f(\phi ,\varphi )-6 a^2 \dot{a} \dot{\zeta} + 6 a \theta  \dot{a}^2 + 6 a \dot{a}^2 - a^3 \dot{\zeta} \dot{\varphi} - a^3 \dot{\theta} \dot{\phi}\,,
\end{equation}
 from which we can derive several interesting theories as shown in the diagram, Fig.~\ref{fig1}. The Noether condition \eqref{symcond}  gives a system of 43 (non-independent) equations for the  Noether vector components
\begin{equation}
\lambda(a,\theta,\phi,\varphi,\zeta,t)\,,\,\,\eta_{a}(a,\theta,\phi,\varphi,\zeta,t)\,,\,\, \eta_{\phi}(a,\theta,\phi,\varphi,\zeta,t)\,,\,\,\eta_{\varphi}(a,\theta,\phi,\varphi,\zeta,t)\,,\,\,\eta_{\theta}(a,\theta,\phi,\varphi,\zeta,t)\,,\,\,\eta_{\zeta}(a,\theta,\phi,\varphi,\zeta,t)\,,
\end{equation}
and the functions
\begin{equation}
h(a,\theta,\phi,\varphi,\zeta,t)\,,\,\,f(\phi,\varphi)\,.
\end{equation}
We can see immediately, from the system, that
\begin{eqnarray}
\lambda(a,\theta,\phi,\varphi,\zeta,t)&=& \lambda (t) \\
\eta_{\phi}(a,\theta,\phi,\varphi,\zeta,t)&=& \eta_{\phi}(a,\phi,\varphi,\zeta,t)\\
h(a,\theta,\phi,\varphi,\zeta,t)&=&h(a,\theta,\phi,\varphi,\zeta)\,.
\end{eqnarray}
The system now reduces to 19 equations that cannot  be easily solved (see Appendix \ref{Appendix} for details). Hence, as we did in the previous sections, we  assume that $h(a,\theta,\phi,\varphi,\zeta) =\mbox{constant} = h$ and after some calculations we end up with the following three equations for $f(\phi,\varphi)$
\begin{eqnarray}\label{eq3}
-f_{\varphi}(\phi ,\varphi ) \left(c_7 \xi  \varphi +c_6 \xi +c_8 \xi -6 c_7 \chi \right)+f_{\phi}(\phi ,\varphi ) \left(-c_5 \xi  \varphi -c_4 \xi +6 c_5 \chi \right)-6 c_7 \chi  \phi  f_{\varphi\phi}(\phi ,\varphi )- \nonumber \\
-6 c_5 \chi  \phi  f_{\phi\phi}(\phi ,\varphi )+c_3 \xi  f(\phi ,\varphi )-c_3+c_{10}-c_{12} = 0\,,\\\label{eq4}
6 \left(c_7-c_3\right) \chi  f_{\varphi}(\phi ,\varphi )+6 \chi  \left(c_7 \varphi +c_6+c_8\right) f_{\varphi\varphi}(\phi ,\varphi )+6 c_5 \chi  f_{\phi}(\phi ,\varphi )+ \nonumber \\
+6 \chi  \left(c_5 \varphi -c_7 \phi +c_4\right) f_{\varphi\phi}(\phi ,\varphi )-6 c_5 \chi  \phi  f_{\phi\phi}(\phi ,\varphi )-c_{12} = 0 \,,\\\label{eq5}
-\left(c_5 \xi +c_3 \chi \right) f_{\phi}(\phi ,\varphi )-c_7 \xi f_{\varphi}(\phi ,\varphi )-6 c_7 \chi  f_{\varphi\varphi}(\phi ,\varphi )+\chi  \left(c_7 \varphi -6 c_5+c_6+c_8\right) f_{\varphi\phi}(\phi ,\varphi )+\nonumber \\
+\chi  \left(c_5 \varphi +c_4\right) f_{\phi\phi}(\phi ,\varphi ) = 0 \,,
\end{eqnarray}
where all the $c$'s are constants coming from the coefficients of the Noether vector. System \eqref{eq3}-\eqref{eq5} can be easily integrated but, depending on the vanishing or not of some constants,  different solutions can be derived. Specifically, we obtain seven different symmetries  described below. The Noether vectors and the function $f$ take the forms:

\begin{enumerate}
\item
\begin{enumerate}
\item
For $c_7 \neq 0 $ and $c_3 \neq 0\,, c_4 \neq \frac{c_5}{c_7}(c_6+c_9)$, we have
\begin{eqnarray}
X=(c_1 t + c_2)\partial_t + \frac{1}{3}(c_1 - c_3)a \partial_{a} + (c_4 + c_5(6 \ln a +  \psi)) \partial_{\phi} + (c_6 + c_7 (6 \ln a  + \varphi )+ c_9) \partial_{\varphi} + c_3 \theta \partial_{\theta}  \nonumber\\
+ ((c_3 - c_7) \zeta  - c_5 \theta + c_8 )\partial_\zeta\,.
\end{eqnarray}
and
\begin{equation}\label{eq6}
f(\phi,\varphi) = \frac{1}{\xi } + \frac{c_{11} \left(c_5 c_6-c_4 c_7+c_5 c_9\right)}{c_3} \exp\left({\frac{c_3}{c_5 c_6-c_4 c_7+c_5 c_9} \left(c_5 \varphi -c_7 \phi \right)} \right)\,.
\end{equation}
\item
For $c_7 \neq 0 $ and $c_3 = 0\,, c_4 = \frac{c_5}{c_7}(c_6+c_9)$, it is 
\begin{equation}
X=(c_1 t + c_2)\partial_t + \frac{c_1}{3} a \partial_{a} + (c_4 + c_5(6 \ln a +  \varphi)) \partial_{\phi} + (c_6 + c_7 (6 \ln a  + \varphi )+ c_9) \partial_{\varphi} + (c_8-c_7 \zeta -c_5 \theta) \partial_{\zeta} \,.
\end{equation}
and
\begin{equation}
f(\phi,\varphi) = c_{11} + F(-c_7 \phi + c_5 \varphi)\,. 
\end{equation}
\end{enumerate}
\item
\begin{enumerate}
\item
\begin{enumerate}
\item
For $c_7 = 0 $ and $c_5 \neq 0$ and $c_3 \neq 0 \,, c_5 \neq - c_6$, it is 
\begin{equation}
X=(c_1 t + c_2)\partial_t + \frac{1}{3}(c_1 - c_3)a \partial_{a} + (c_4 + c_5(6 \ln a +  \varphi)) \partial_{\phi} + (c_6 + c_9) \partial_{\varphi} + (c_{10}+c_3 \theta )\partial_{\theta}  + (c_3  \zeta  - c_5 \theta + c_8 )\partial_\zeta\,,
\end{equation}
and
\begin{equation}
f(\phi,\varphi) = \frac{c_3-c_{10}}{\xi c_3 } + c_{11} e^{\frac{c_3}{c_6 c_9} \varphi}\,. 
\end{equation}
\item
For $c_7 = 0 $ and $c_5 \neq 0$ and $c_3  = 0 \,, c_5 = - c_6$, it is
\begin{equation}
X=(c_1 t + c_2)\partial_t + \frac{c_1}{3}a \partial_{a} + (c_4 + c_5(6 \ln a +  \varphi)) \partial_{\phi}  + (c_8 - c_5 \theta )\partial_\zeta\,.
\end{equation}
and
\begin{equation}
f(\phi,\varphi) =  c_{11} + F (\varphi)\,. 
\end{equation}
\end{enumerate}
\item
\begin{enumerate}
\item
For $c_7 = 0 $ and $c_5 = 0$ and $c_3 \neq 0 \,, c_4 \neq 0$, it is 
\begin{equation}
X=(c_1 t + c_2)\partial_t + \frac{1}{3}(c_1 - c_3)a \partial_{a} + c_4\partial_{\phi} + (c_6 + c_9) \partial_{\varphi} + (c_{10}+c_3 \theta )\partial_{\theta}  + (c_8 + c_3  \zeta )\partial_\zeta\,,
\end{equation}
and
\begin{equation}
f(\phi,\varphi) = \frac{c_3-c_{10}}{\xi c_3 } + F(-\frac{c_6+c_9}{c_4}\phi+\varphi) e^{\frac{c_3}{c_4} \phi}\,. 
\end{equation}
\item
\begin{enumerate}
\item
For $c_7 = 0 $ and $c_5 = 0$ and $c_3 = 0 \,, c_4 = 0$ and $c_6 \neq -c_7$, it is 
\begin{equation}
X=(c_1 t + c_2)\partial_t + \frac{c_1}{3}a \partial_{a}  + (c_6 + c_9) \partial_{\varphi} + c_{10}\partial_{\theta}  + c_8\partial_\zeta\,,
\end{equation}
and
\begin{equation}
f(\phi,\varphi) = \frac{c_{10}}{(c_6+c_9)\xi }\varphi + F(\phi)\,. 
\end{equation}
\item
For $c_7 = 0 $ and $c_5 = 0$ and $c_3 = 0 \,, c_4 = 0$ and $c_6 = -c_7$, it is
\begin{equation}
X=(c_1 t + c_2)\partial_t + \frac{c_1}{3}a \partial_{a}  + c_8 \partial_\zeta\,,
\end{equation}
and the equations are satisfied for any $f$.
\end{enumerate}
\end{enumerate}
\end{enumerate}
\end{enumerate}
Clearly, each of these symmetries specify a different Lagrangian and then a different dynamics. As discussed in the Appendix \ref{Appendix}, the fact that several symmetries exist for the same symmetry condition \eqref{symcond} is due to the fact that such a condition consists in a system of non-linear partial differential equations which have no unique general solution.

\subsection{Cosmological Solutions}
Let us now find  cosmological solutions for the generalized Lagrangian \eqref{genlagra}. In principle, it is possible to find out  cosmological solutions for each of the above cases depending on the coupling functions. Due to the physical importance of the exponential couplings, we will  present  cosmological solutions for the  coupling function given by (\ref{eq6}). However, the procedure for the other cases is the same.

In the case (\ref{eq6}), we have the constraint given by the integration constants, that is  $c_7 \neq 0 \,, c_3 \neq 0\,, c_4 \neq \frac{c_5}{c_7}(c_6+c_9)$.  Hence, the Euler-Lagrange equations obtained by \eqref{genlagra}, together with the energy condition, give a system of six differential equations for $a(t)\,,\phi(t)\,,\varphi(t)\,,\theta(t)\,\text{and}\,\zeta(t)$. \\
Assuming that  the scale factor of the universe behaves as de-Sitter $a(t)= e^{H_0 t}$, it is possible to find different kind of solutions depending on different cases for the constants. In all of these cases, the final cosmological solutions are almost the same. A general solution that one can easily find is 
\begin{gather}
a(t) = e^{H_0 t}\,, \phi (t) = -2 H_0 t\,, \theta (t)  = \frac{1}{3} e^{-3 H_{0} t} \left(-\frac{18 c_{11} c_7 \chi ^2 (c_7-3 c_5)^2 \exp \left(\frac{H_{0} t (3 c_5-2 c_7) (\xi +3 \chi )}{\chi  (3 c_5-c_7)}\right)}{(3 c_5-2 c_7) (3 c_5 \xi -2 c_7 \xi -3 c_7 \chi )}-\frac{\theta_{1}}{H_{0}}\right)\,, \\
\varphi (t) = -6 H_0 t \,, \zeta(t) = \frac{1}{3} e^{-3 H_{0} t} \left(\frac{18 c_{11} c_5 \chi ^2 (c_7-3 c_5)^2 \exp \left(\frac{H_{0} t (3 c_5-2 c_7) (\xi +3 \chi )}{\chi  (3 c_5-c_7)}\right)}{(3 c_5-2 c_7) (3 c_5 \xi -2 c_7 \xi -3 c_7 \chi )}-\frac{\zeta_{1}}{H_{0}}\right)+\zeta_{0}\,,
\end{gather}
and the coupling function $f$ becomes
\begin{equation}
f(\phi,\varphi) = \frac{1}{\xi }-\frac{2 c_{11} \chi  (c_7-3 c_5)^2 }{3 c_5 \xi -2 c_7 \xi -3 c_7 \chi }\exp \left(-\frac{(3 c_5 \xi -2 c_7 \xi -3 c_7 \chi ) (c_5 \varphi -c_7 \phi )}{2 \chi  (c_7-3 c_5)^2}\right)\,,
\end{equation}
where $\theta_1\,,\zeta_1$ and $\zeta_2$ are integration constants and we need to set ${\displaystyle c_3=-\frac{(3 c_5 \xi -2 c_7 \xi -3 c_7 \chi ) (-c_4 c_7+c_5 c_6+c_5 c_9)}{2 \chi  (3 c_5-c_7)^2}}$. 
Apart from de-Sitter solutions, the system admits also power-law solutions. For example, by setting ${\displaystyle c_3=\frac{\left(9 p^2-9 p+2\right) (-c_4 c_7+c_5 c_6+c_5 c_9)}{6 p (3 c_5 p-c_5-c_7 p)}}$, we get the following solutions
\begin{gather}
a(t)=t^p\,,\phi (t)=\frac{6 p^2 \ln (t-3 p t)}{1-3 p}\,,\varphi (t)=-6 p \ln t\,,\theta (t)=\frac{6 c_{11} c_7 p t^{2-3 p} (1-3 p)^{\frac{c_7 (2-3 p) p}{-3 c_5 p+c_5+c_7 p}} (p (\xi +3 \chi )-\chi )}{3 p-2}+\frac{\theta_{1} t^{1-3 p}}{1-3 p}\,, \\
\zeta (t)-\frac{6 c_{11} c_5 p t^{2-3 p} (1-3 p)^{-\frac{c_7 p (3 p-2)}{-3 c_5 p+c_5+c_7 p}} (p (\xi +3 \chi )-\chi )}{3 p-2}+\zeta_{0}+\frac{\zeta_{1} t^{1-3 p}}{1-3 p}\,,
\end{gather}
and the coupling function $f$ becomes
\begin{equation}
f(\phi,\varphi) = \frac{1}{\xi }-\frac{6 c_{11} p (-3 c_5 p+c_5+c_7 p) }{9 p^2-9 p+2}\exp \left(-\frac{\left(9 p^2-9 p+2\right) (c_5 \varphi -c_7 \phi )}{6 p (-3 c_5 p+c_5+c_7 p)}\right)\,.
\end{equation}
The above procedure can be iterated for all the above couplings. We stress again the important fact that such couplings are not arbitrarily given but result from the existence of the symmetries.

\section{Discussion and Conclusions}
\label{conclusions}			
Motivated by an increasing amount of studies related to non-local theories,  here we proposed a new generalized non-local theory of gravity including curvature and teleparallel terms. These kind of theories were introduced motivated by loop quantum effects and they have attracted a lot of interest since some of them are renormalizable \cite{ModestoQ1}. Under suitable limits, the  general action that we proposed can represent either curvature non-local theories with $Rf(\square^{-1}R)$ based on \cite{Deser:2007jk} or teleparallel nonlocal theories $Tf(\square^{-1}T)$ based on \cite{our}. Since the theory is highly non-linear, it is possible to introduce  four auxiliary scalar fields in order to rewrite the action in an easier way. Then, for a flat FRW cosmology, using the Noether Symmetry Approach,  the coupling functions can be selected directly from the symmetries for the various models derived from the general theory. It is obvious that the theory \eqref{ourtheory} can give several models, depending on the values of the constants $\xi$ and $\chi$ and on the form of the distortion function. We prove that, in most physically  interesting cases, the only forms of the distortion function selected by the Noether Symmetries, are the exponential and the linear ones. According to the literature 
\cite{Nojiri:2007uq,Jhingan:2008ym}, this is an important result, because, up to now, these kinds of couplings were chosen by hand in order to find cosmological solutions while, in our case, they come out from a first principle. In addition, there is a specific class of exponentials non-local gravity models which are renormalizable (see \cite{Mod3,ModestoQ2}). This means that, 	the Noether Symmetries dictate the form of the action and choose exponential  form for the distortion function.   As discussed in \cite{NoetherQC}, the existence of Noether's symmetries is a selection criterion for physically motivated models.
Finally, from models selected by symmetries, it is easy to find cosmological solutions like  de-Sitter and power-law ones. The integrability of dynamics is guaranteed by the existence of first integrals. In forthcoming studies, the cosmological analysis will be improved in view of observational data.

\begin{acknowledgments}
 SB is supported by the Comisi{\'o}n Nacional de Investigaci{\'o}n Cient{\'{\i}}fica y Tecnol{\'o}gica (Becas Chile Grant No.~72150066).
SC and KFD are supported in part by the INFN sezione di Napoli, {\it iniziative specifiche} TEONGRAV and QGSKY.
The  article is also based upon work from COST action CA15117 (CANTATA), 
supported by COST (European Cooperation in Science and Technology).  
\end{acknowledgments}

\appendix 

\section{The existence conditions for Noether's symmetry }
\label{Appendix}
The vector equation  \eqref{symcond}
\begin{equation}
X^{[1]}\mathcal{L}+\mathcal{L}\frac{d\lambda}{dt} = 
\frac{dh}{dt}\,,
\end{equation}
that we rewrite here for  simplicity, gives  the existence conditions for the Noether symmetries.  This means that  the Noether vector components $\eta^i$ and the functions $\lambda$ and $h$ have to be selected.  If some of these functions are different from zero, the symmetry exists and, vice versa, selects the class of possible Lagrangians $ \mathcal{L}$ compatible with it. 
These conditions  constitute a system of partial differential equations derived by equating to zero the coefficients of time derivatives in Eq. \eqref{symcond}.The result of this system are the non-trivial functions $\eta^i$, $\lambda$, and $h$. The  number of differential equations depends on the dimension of the configuration space ${\mathcal Q}$.
For a detailed discussion on the method,  see \cite{cimento, Paliathanasis:2015mxa}.

In the present case, considering  the general  Lagrangian \eqref{genlagra}, the  full Noether conditions are  43 differential equations. However, there are 24 differential equations that can be easily solved giving us the  functions
\begin{eqnarray}
\label{functionsN}
\lambda=\lambda(t)\,,\quad h=h(a,\theta,\phi,\psi,\zeta)\,,\quad \eta_{\phi}=\eta_{\phi}(a,\phi,\psi,\zeta,t)\,.
\end{eqnarray}
The remaining 19 differential equations are the followings:

\begin{eqnarray}
h_{,\theta}+a^3\eta_{\phi,t}=0\,, \\
 h_{,\psi}+a^2 \left(a \eta_{\zeta,t}-6 \chi f_{,\psi} \eta_{a,t}\right)=0\,,\\
 h_{,\phi}+a^2 \left(a \eta_{\theta,t}-6 \chi  f_{,\phi}\eta_{a,t} \right)=0\,,\\
h_{,\zeta}+a^2 \left(6\eta_{a,t} +a\eta_{\psi,t} \right)=0\,,\\
 a \left(\eta_{\phi,\psi}+\eta_{\zeta,\theta}\right)-6 \chi f_{,\psi} \eta_{a,\theta}=0\,,\\
  a \left(\eta_{\phi,\zeta}+\eta_{\psi,\theta}\right)+6 \eta_{a,\theta}=0\,,\\
6 \chi f_{,\phi} \eta_{a,\phi} -a \eta_{\theta,\phi} =0\,,\\
  6 \chi f_{,\psi}\eta_{a,\psi}-a\eta_{\zeta,\psi}=0\,,\\
   6 \eta_{a,\zeta}+a \eta_{\psi,\zeta}=0\,,\\
h_{,a}-6 a^2 \left(\chi f_{,\phi} \eta_{\phi,t} +\chi f_{,\psi} \eta_{\psi,t} -\eta_{\zeta,t}\right)+12 a (\xi  f-\theta-1) \eta_{a,t}=0\,,\\
6 a\left(\chi f_{,\psi} \eta_{\psi,\theta}-\eta_{\zeta,\theta}\right)+12 (\theta+1-\xi  f) \eta_{a,\theta}-a^2 \eta_{\phi,a}=0\,,\\
6 \chi (f_{,\phi} \eta_{a,\psi}+ f_{,\psi}  \eta_{a,\phi})-a \left(\eta_{\theta,\psi}+\eta_{\zeta,\phi}\right)=0\,,  \\
6 (\chi f_{,\phi}\eta_{a,\zeta}-\eta_{a,\phi})-a \left(\eta_{\psi,\phi}+\eta_{\theta,\zeta}\right)=0\,,\\
6 \chi f_{,\phi} \eta_{a,\theta}-a \left(\eta_{\phi,\phi}+\eta_{\theta,\theta}-\lambda_{,t}\right)-3\eta_{a}=0\,,\\
6 \chi f_{,\psi}\eta_{a,\zeta} -6\eta_{a,\psi}-3 \eta_{a}-a \left(\eta_{\psi,\psi}+\eta_{\zeta,\zeta}-\lambda_{,t}\right)=0\,,\\
6 a \left(\chi f_{,\phi} \eta_{\phi,\phi}+\chi f_{,\phi\phi} \eta_{\phi}+\chi f_{,\phi}\eta_{a,a} +\chi f_{,\psi}\eta_{\psi,\phi}+\chi f_{,\phi\psi}\eta_{\psi}-\chi  \lambda_{,t}f_{,\phi}-\eta_{\zeta,\phi}\right)+12 \chi f_{,\phi}\eta_{a}+12 (\theta +1-\xi  f) \eta_{a,\phi}\nonumber\\
-a^2 \eta_{\theta,a}=0\,,\\
6 a \left(\chi f_{,\phi} \eta_{\phi,\zeta}+\chi f_{,\psi} \eta_{\psi,\zeta}-\eta_{a,a}-\eta_{\zeta,\zeta}+\lambda_{t}\right)+12 (\theta +1-\xi  f) \eta_{a,\zeta}-12 \eta_{a} -a^2 \eta_{\psi,a}=0\,,\\
6 a \left(\chi f_{,\phi}\eta_{\phi,\psi}+\chi f_{,\phi\psi}\eta_{\phi}+\chi f_{,\psi} \left(\eta_{a,a}+\eta_{\psi,\psi}-\lambda_{t}\right)+\chi f_{,\psi\psi} \eta_{\psi} -\eta_{\zeta,\psi}\right)+12 \chi f_{,\psi}\eta_{a}+12 (\theta +1-\xi  f) \eta_{a,\psi}\nonumber\\
-a^2 \eta_{\zeta,a}=0\,,\\
a \left(\chi  a f_{,\phi}\eta_{\phi,a}-\xi f_{,\phi}\eta_{\phi}+\chi  a f_{,\psi}\eta_{\psi,a}-\xi f_{,\psi}\eta_{\psi}-2 \xi  f \eta_{a,a}+\xi  \lambda_{,t}f+2 \theta \eta_{a,a}+2  \eta_{a,a}+\eta_{\theta}-a \eta_{\zeta,a}-  (\theta+1)\lambda_{t}\right)\nonumber\\
+(\theta+1-\xi  f)\eta_{a}=0\,.
\end{eqnarray}
where commas denote partial derivatives. The solutions of these equations are the functions \eqref{functionsN}. Clearly, being a system of non-linear partial differential equations, the solution is not unique. This means that several Noether symmetries can be selected according to the different functions \eqref{functionsN}.


\begin{thebibliography}{99}
\bibitem{Capozziello:2011et}
  S.~Capozziello and M.~De Laurentis,
  Phys.\ Rept.\  {\bf 509} (2011) 167
  doi:10.1016/j.physrep.2011.09.003
  [arXiv:1108.6266 [gr-qc]].
  
  \bibitem{reportsergei}
 S.D. Odintsov and S. Nojiri, Phys. Rept. {\bf 505}, 59 (2011).
  doi: 10.1016/j.physrep.2011.04.001.
  [arXiv:1011.0544 [gr-qc]].

\bibitem{Clifton:2011jh}
  T.~Clifton, P.~G.~Ferreira, A.~Padilla and C.~Skordis,
  Phys.\ Rept.\  {\bf 513} (2012) 1
  doi:10.1016/j.physrep.2012.01.001
  [arXiv:1106.2476 [astro-ph.CO]].
  
 
\bibitem{Deser:2007jk}
  S.~Deser and R.~P.~Woodard,
  Phys.\ Rev.\ Lett.\  {\bf 99} (2007) 111301
  doi:10.1103/PhysRevLett.99.111301
  [arXiv:0706.2151 [astro-ph]].
  
\bibitem{Donoghue:1994dn}
  J.~F.~Donoghue,
  Phys.\ Rev.\ D {\bf 50} (1994) 3874
  doi:10.1103/PhysRevD.50.3874
  [gr-qc/9405057].
  
\bibitem{Giddings:2006sj}
  S.~B.~Giddings,
  Phys.\ Rev.\ D {\bf 74} (2006) 106005
  doi:10.1103/PhysRevD.74.106005
  [hep-th/0605196].
  
  \bibitem{univ4} S. Das and E. C. Vagenas, Phys. Rev. Lett. 101, 221301 (2008)  


\bibitem{modesto1}
 L. Modesto, I.L. Shapiro,   Phys.Lett. B { 755}, 279  (2016).

\bibitem{modesto2}  L. Modesto,  Nucl.Phys. B { 909}, 584  (2016). 

\bibitem{st1}
G.  Calcagni and  L. Modesto, J. Phys. A: Math. Theor. 47, 355402 (2014)

\bibitem{st2}
G.  Calcagni and  G.  Nardelli, Phys. Rev. D82, 123518 (2010)

\bibitem{loop} S. A. Major and M. D. Seifert, Class. Quant. Grav. 19, 2211  (2002)  

\bibitem{jm} L.  Modesto, J. W. Moffat and P. Nicolini,  Phys. Lett. B 695, 397 (2011)

\bibitem{Mod1}
L.~Modesto and L.~Rachwal,
  Nucl.\ Phys.\ B {\bf 889}, 228 (2014)
  [arXiv:1407.8036 [hep-th]].

\bibitem{Mod2}
 L.~Modesto and L.~Rachwal,
  arXiv:1605.04173 [hep-th].
  

\bibitem{Mod3}
L.~Modesto and L.~Rachwal,
Int.\ J.\ Mod.\ Phys.\ D {\bf 26}, 1730020 (2017).

\bibitem{Mod4}
A.~S.~Koshelev, L.~Modesto, L.~Rachwal and A.~A.~Starobinsky,
  JHEP {\bf 1611}, 067 (2016)
  [arXiv:1604.03127 [hep-th]].

\bibitem{Mod5}
L.~Modesto and L.~Rachwal,
  Nucl.\ Phys.\ B {\bf 900}, 147 (2015)
  [arXiv:1503.00261 [hep-th]].
  
\bibitem{Arefeva:2007wvo}
  I.~Y.~Aref'eva, L.~V.~Joukovskaya and S.~Y.~Vernov,
  JHEP {\bf 0707} (2007) 087
  doi:10.1088/1126-6708/2007/07/087
  [hep-th/0701184].
  
  
\bibitem{Arefeva:2007xdy}
  I.~Y.~Aref'eva, L.~V.~Joukovskaya and S.~Y.~Vernov,
  J.\ Phys.\ A {\bf 41} (2008) 304003
  doi:10.1088/1751-8113/41/30/304003
  [arXiv:0711.1364 [hep-th]].
  
\bibitem{Barnaby:2008fk}
  N.~Barnaby and J.~M.~Cline,
  JCAP {\bf 0806} (2008) 030
  doi:10.1088/1475-7516/2008/06/030
  [arXiv:0802.3218 [hep-th]].
  

  
\bibitem{Jhingan:2008ym}
  S.~Jhingan, S.~Nojiri, S.~D.~Odintsov, M.~Sami, I.~Thongkool and S.~Zerbini,
  Phys.\ Lett.\ B {\bf 663} (2008) 424
  doi:10.1016/j.physletb.2008.04.054
  [arXiv:0803.2613 [hep-th]].
  
  
\bibitem{Deser:2013uya}
  S.~Deser and R.~P.~Woodard,
  JCAP {\bf 1311} (2013) 036
  doi:10.1088/1475-7516/2013/11/036
  [arXiv:1307.6639 [astro-ph.CO]].
  
\bibitem{Deffayet:2009ca}
  C.~Deffayet and R.~P.~Woodard,
  JCAP {\bf 0908} (2009) 023
  doi:10.1088/1475-7516/2009/08/023
  [arXiv:0904.0961 [gr-qc]].

\bibitem{Koivisto:2008xfa}
  T.~Koivisto,
  Phys.\ Rev.\ D {\bf 77} (2008) 123513
  doi:10.1103/PhysRevD.77.123513
  [arXiv:0803.3399 [gr-qc]].
  
\bibitem{Koivisto:2008dh}
  T.~S.~Koivisto,
  Phys.\ Rev.\ D {\bf 78} (2008) 123505
  doi:10.1103/PhysRevD.78.123505
  [arXiv:0807.3778 [gr-qc]].



\bibitem{Barvinsky:2014lja}
  A.~O.~Barvinsky,
  Mod.\ Phys.\ Lett.\ A {\bf 30} (2015) no.03n04,  1540003
  doi:10.1142/S0217732315400039
  [arXiv:1408.6112 [hep-th]].

\bibitem{Pereira}
R. Aldrovandi and J. G. Pereira, {\it Teleparallel Gravity: An Introduction}, Springer, Dordrecth, (2012), "Teleparallel Gravity" at http://www.ift.unesp.br/users/jpereira/tele.pdf


\bibitem{Cai:2015emx}
  Y.~F.~Cai, S.~Capozziello, M.~De Laurentis and E.~N.~Saridakis,
  Rept.\ Prog.\ Phys.\  {\bf 79} (2016) no.10,  106901
  doi:10.1088/0034-4885/79/10/106901
  [arXiv:1511.07586 [gr-qc]].

\bibitem{manos}
G. Kofinas and E. N. Saridakis, Phys. Rev. D
{\bf 90}, 084044 (2014).

\bibitem{Capozziello:2016eaz}
  S.~Capozziello, M.~De Laurentis and K.~F.~Dialektopoulos,
  Eur.\ Phys.\ J.\ C {\bf 76} (2016) no.11,  629
  doi:10.1140/epjc/s10052-016-4491-0
  [arXiv:1609.09289 [gr-qc]].
  
  \bibitem{Paliathanasis:2016vsw}
  A.~Paliathanasis, J.~D.~Barrow and P.~G.~L.~Leach,
  Phys.\ Rev.\ D {\bf 94} (2016) no.2,  023525
  doi:10.1103/PhysRevD.94.023525
  [arXiv:1606.00659 [gr-qc]].
  
\bibitem{Bahamonde:2016kba}
  S.~Bahamonde and C.~G.~B\"ohmer,
  Eur. Phys. J. C {\bf 76} (2016) no.10,  578
  doi:10.1140/epjc/s10052-016-4419-8
  [arXiv:1606.05557 [gr-qc]].
  
  
\bibitem{Chen:2010va}
  S.~H.~Chen, J.~B.~Dent, S.~Dutta and E.~N.~Saridakis,
  Phys.\ Rev.\ D {\bf 83} (2011) 023508
  doi:10.1103/PhysRevD.83.023508
  [arXiv:1008.1250 [astro-ph.CO]].


\bibitem{Li:2011wu}
  B.~Li, T.~P.~Sotiriou and J.~D.~Barrow,
  Phys.\ Rev.\ D {\bf 83} (2011) 104017
  doi:10.1103/PhysRevD.83.104017
  [arXiv:1103.2786 [astro-ph.CO]].
      
  \bibitem{our} 
  S.~Bahamonde, S.~Capozziello, M.~Faizal and R.~C.~Nunes,
  Eur.\ Phys.\ J.\ C {\bf 77} (2017) no.9,  628
  doi:10.1140/epjc/s10052-017-5210-1
  [arXiv:1709.02692 [gr-qc]].
  
  
  
  
  
  
 \bibitem{cimento} 
  S. Capozziello, R. De Ritis, C. Rubano, P. Scudellaro,  {\it Riv.Nuovo Cim.} {\bf 19(4)}, 1, (1996).  
  
\bibitem{Bahamonde:2016grb}
S.~Bahamonde and S.~Capozziello,
Eur.\ Phys.\ J.\ C {\bf 77} (2017) no.2,  107
[arXiv:1612.01299 [gr-qc]].

  
  \bibitem{NoetherQC}
S. Capozziello, M. De Laurentis, S.D. Odintsov, 
 Eur.\ Phys.\ J.\ C {\bf 72} (2016),  2068
 doi: 10.1140/epjc/s10052-012-2068-0.
 
\bibitem{Nojiri:2007uq}
  S.~Nojiri and S.~D.~Odintsov,
  Phys.\ Lett.\ B {\bf 659} (2008) 821
  doi:10.1016/j.physletb.2007.12.001
  [arXiv:0708.0924 [hep-th]].

\bibitem{Paliathanasis:2015mxa}
A. Borowiec,  S. Capozziello, M. De Laurentis, F. S. N. Lobo,
A. Paliathanasis, M. Paolella,  and A. Wojnar,
Phys. Rev.  D {\bf 91}, 023517 (2015)
doi: 10.1103/PhysRevD.91.023517.
  

\bibitem{Basilakos}
S. Basilakos, S. Capozziello,  M. De Laurentis,  A. Paliathanasis,  and M. Tsamparlis,  Phys. Rev. D {\bf 88}, 103526 (2013)
doi: 10.1103/PhysRevD.88.103526.


\bibitem{ModestoQ2}
L. Modesto and  L. Rachwal, arXiv:1605.04173 [hep-th] (2016).






\end{thebibliography}
\end{document}